\documentclass[a4paper,12pt]{article}
\usepackage{epsfig}
\makeatletter
\@addtoreset{equation}{section}
\makeatother
\def\theequation{\thesection.\arabic{equation}}
\def\appendix{\setcounter{section}{0}
\def\thesection{Appendix \Alph{section}}
\def\theequation{\Alph{section}.\arabic{equation}}}
\newcommand{\PMS}{\rm PMS}
\newcommand{\MS}{\rm MS}
\newcommand{\eff}{\rm eff}
\newcommand{\an}{\rm an}
\newcommand{\TV}{\rm PT}
\newcommand{\ATV}{\rm APT}
\newcommand{\pole}{\rm pole}
\newcommand{\appr}{\rm approx}
\newcommand{\cut}{\rm cut}
\newcommand{\Ima}{\rm Im}
\newcommand{\Req}{\rm Re}
\newcommand{\rres}{ res}
\newcommand{\sing}{ sing}

\begin{document}
\thispagestyle{empty}

\noindent\phantom{}\hspace{10cm}{\sf Dedicated to the memory of}\\
\noindent\phantom{}\hspace{10cm}{\sf Nikolai Nikolaevich Bogoliubov}

\noindent\phantom{}\vspace{1.0cm}
\begin{center}
{\Large ANALYTIC APPROACH IN QUANTUM CHROMODYNAMICS}\\
\vspace{1.3cm}
{\large I.L.~Solovtsov and D.V.~Shirkov}\\
\vspace{0.7cm}
{\it Bogoliubov Laboratory of Theoretical Physics, \\
Joint Institute for Nuclear Research, Dubna, 141980 Russia}
\end{center}

\vspace{1.3cm}
\begin{abstract} We investigate a new ``renormalization invariant analytic
formulation'' of calculations in quantum chromodynamics, where the
renormalization group summation is correlated with the analyticity with
respect to the square of the transferred momentum~$Q^2$.  The expressions
for the invariant charge and matrix elements are then modified such that the
unphysical singularities of the ghost pole type do not appear at all, being
by construction compensated by additional nonperturbative contributions.
Using the new scheme, we show that the results of calculations for a number
of physical processes are stable with respect to higher-loop effects and the
choice of the renormalization prescription.

Having in mind applications of the new formulation to inelastic
lepton--nucleon scattering processes, we analyze the corresponding structure
functions starting from the general principles of the theory expressed by
the Jost--Lehmann--Dyson integral representation.  We use a nonstandard
scaling variable that leads to modified moments of the structure functions
possessing K\"all\'en--Lehmann analytic properties with respect to~$Q^2$. We
find the relation between these ``modified analytic moments'' and the
operator product expansion.
\end{abstract}

\newpage
\phantom{}\vspace{0.7cm}
\noindent\phantom{}\hspace{10.1cm}{\it Take care of the Principles, and the}\\
\noindent\phantom{}\hspace{10.1cm}{\it Principles shall take care of you.}

\bigskip
\vspace{1.7cm}

{\sl Scientific achievements of Nikolai Nikolaevich Bogoliubov are
characterized by a unique combination of determination in solving
concrete scientific problems and a high level of mathematical culture.
He could find the shortest path to a physical result using most
general principles of the theory.

The renormalization-invariant analytic approach to quantum
chromodynamics exposed here and its most recent applications are based
on the works~\cite{1,2,3,4} by Bogoliubov with his closest collaborators.
A characteristic feature of these investigations is their strong
relation with the fundamental quantum physics principles\/}.

\vspace{2.9cm}
\tableofcontents

\newpage\vspace{1.1cm}
\section{Introduction}

An intrinsic ingredient of modern quantum field theory (QFT) is the
renormalization group (RG) method proposed in the
mid-fifties~\cite{1,2}.  The role of this method is particularly
important in the cases where the interaction is not weak, for example,
in quantum chromodynamics (QCD).  Hardly any hadronic process
investigated in the QCD framework can be analyzed without using the
renormalization group.  It is well known that directly solving the RG
equation for the invariant charge leads to unphysical singularities,
for example, to the ghost pole in the one-loop approximation.
Taking next loop corrections into account does not alter the essence,
and leads only to additional branch cuts.  The existence of such
singularities contradicts the general principles of local~QFT.

As early as in the late-fifties, N.\,N.~Bogoliubov and
collaborators~\cite{3} proposed a resolution of this problem in the
context of quantum electrodynamics (QED) by unifying the RG method
with the requirement of analyticity with respect to~$Q^2$, which in
turn followed from the known K\"all\'en--Lehmann representation
expressing the basic principles of local QFT~\cite{5} [see
Eq.~(\ref{2.1}) below].

The invariant QED charge $\bar{\alpha}(Q^2)$ (also referred to as the
``invariant, or running coupling constant"\footnote{In view of semantical
absurdity of the last term, we use the expression {\sl invariant coupling
function\/} or {\sl invariant coupling}.}) is proportional to the transverse
amplitude of the full photon propagator, which satisfies the spectral
K\"all\'en--Lehmann representation corresponding to the {\sl analyticity in
the complex~$Q^2$ plane cut along the negative part\footnote{We use the
notation $Q^2=-q^2$, hence the Euclidean region corresponds to
positive~$Q^2$.} of the real axis}.  According to~\cite{3}, the analytic
invariant charge can be reconstructed via the K\"all\'en--Lehmann
representation, in which the relevant spectral density is defined as the
imaginary part of the invariant charge determined by the RG method in the
Euclidean region and analytically continued to the domain where
$\Req\,Q^2<0$. The explicit one-loop (and implicit two-loop) expression
obtained in~\cite{3} for the analytic coupling in QED has the following
important {\sl properties}:

-- {\sl the ghost pole is absent};

--  {\sl as a function of $\alpha$, this expression has an essential
    singularity in the neighborhood of $\alpha=0$ of the form
    $\exp(-3\pi/\alpha)$};

-- {\sl for real positive~$\alpha$, it admits an expansion in powers
    of~$\alpha$ that coincides with the perturbative expansion};

-- {\sl it has a finite ultraviolet {\rm(}UV\/{\rm)} limit equal to
    $3\pi$, which is independent of the experimental
    value~$\alpha\simeq1/137$}.

In~\cite{6,7}, the idea to combine the renormalization invariance and
the $Q^2$-analyticity in QCD led to uncovering new important
properties of the analytic coupling.  These properties include the
existence of an infrared fixed point of $\bar{\alpha}_{\an}(Q^2)$,
which proves to be universal in the sense that its value
$\alpha=4\pi/\beta_0$ is already determined by the one-loop
contribution (i.e., remains unchanged by the multiloop corrections and
is therefore scheme-invariant).  It is also independent of the
experimentally determined QCD parameter~$\Lambda$, and the set of
curves $\bar{\alpha}_{\an}(Q^2/\Lambda^2)$ corresponding to different
values of~$\Lambda$ is a bundle with the common point
$\bar{\alpha}_{\an}(0)=4\pi/\beta_0$.  Thus, the analytic approach
leads to essential modifications of the infrared (IR) behavior of the
perturbative invariant coupling.  We give the approximate formulas
that are useful in the two-loop approximation and also discuss some
phenomenological applications of the analytic approach.\footnote{The
works~[8-16]
are devoted to the development and applications of the analytic approach.}

This work can be conventionally divided into three parts.  In the
first one (Sec.~2), which is a review of our publications over the
last two years, the analytic invariant approach is formulated in
general and is explained in detail in application to the analytic
coupling ``constant.''  In the second part, which is also a review
(Sec.~3), we formulate the ``analytic perturbation'' theory for
physical quantities expressed through the two-point objects of the
type of the Adler $D(Q^2)$ function, whose properties can be related
to the K\"all\'en--Lehmann representation; we also discuss there the
problems of scheme and loop dependence.

In the third part (Sec.~4), we finally consider the structure
functions (formfactors) parametrizing the inelastic lepton--hadron
scattering cross-section.  To relate them to analytic functions
of~$Q^2$, we start with the Jost--Lehmann--Dyson integral
representation.  Using the results of Bogoliubov, Vladimirov, and
Tavkhelidze~\cite{4}, we adduce the arguments in favor of the
introduction of a special scaling variable such that the moments of
the structure functions with respect to this variable admit a
K\"all\'en--Lehmann representation.  This allows us to apply the
analyticization procedure to these moments.  We also consider the
relation of the analytic moments with the operator product expansion.

\sloppy
\section{An invariant analytic formulation of QCD}
\label{sec2}
\fussy

In this section, we formulate the method of constructing the analytic
invariant charge and consider its main properties.

\subsection {The renormalization group and analyticity}

We start with two remarks.  It is known that the invariant QCD charge
$\bar{\alpha}_s(Q^2)$ is defined via the product of propagators and the
special vertex functions, which gives rise to the problem of whether the
spectral representation can be used for this product.  This problem was
studied in~\cite{17}, where it was shown that the invariant coupling can be
written in the form of a spectral integral.  In the general case, in
addition, the evolution of $\bar{\alpha}_s(Q^2)$ is related to the
``running'' gauge parameter. For simplicity, we use the standard
$\overline{\rm MS}$-scheme, where the gauge does not affect the invariant
charge.\footnote{A similar situation occurs in the MOM-scheme in the
transverse gauge or in the MOM-scheme when applying a special
renormalization~\cite{18}.}

We write the spectral representation for the invariant coupling
$a(Q^2)=\alpha_s(Q^2)/(4\pi)$ as
\begin{equation}
\bar{a}_{\an}(Q^2)=\frac{1}{\pi}\int_0^{\infty}
d\sigma\frac{\rho(\sigma,a)}{\sigma+Q^2-i\epsilon}.
\label{2.1}
\end{equation}
In the perturbation theory summed up in accordance with the
renormalization group, the spectral density $\rho(\sigma,a)$ decreases
as $1/\log^2\sigma$, which allows us to write the spectral
representation without subtractions.

           \begin{figure}[ht]
\centerline{ \epsfig{file=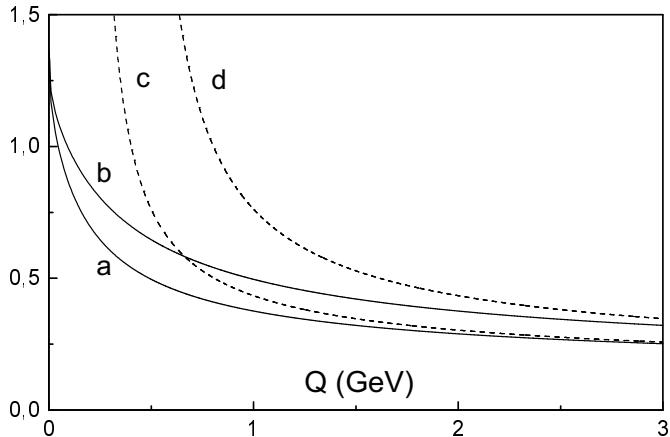,width=10.7cm}} \caption{   {\sl  The
behavior of the one-loop analytic coupling
$\bar{\alpha}_{\an}(Q^2)$: {\sl a}~for $\Lambda=200\,$MeV, {\sl b}~for
$\Lambda=400\,$MeV.  The curves~{\sl c} and~{\sl d} correspond to
perturbation theory for the same value of~$\Lambda$.
           }       }
         \label{shir1}
         \end{figure}

In the leading logarithmic approximation, the invariant coupling has
the form
\begin{equation}
\bar{a}^{(1)}(Q^2)=\frac{a}{1+a\beta_0\log(Q^2/\mu^2)}=
\frac{1}{\beta_0\log(Q^2/\Lambda^2)},
\label{2.2}
\end{equation}
where $\beta_0=11-2f/3$ is the one-loop $\beta$-function
coefficient with $f$ active quarks and the QCD scaling parameter
is~$\Lambda=\mu\exp[-1/2a_{\mu}\beta_0]$.  The corresponding spectral
density reads as
\begin{equation}
\frac{a^2\beta_0\pi}
{\bigl[1+a\beta_0\log(\sigma/\mu^2)\bigr]^2+[a\beta_0\pi]^2}=
\frac{1}{\beta_0}\frac{\pi}{\log^2(\sigma^2/\Lambda^2)+\pi^2}=
\rho^{(1)}(\sigma,a).
\label{2.3}
\end{equation}
Inserting this into spectral integral~(\ref{2.1}) gives the
one-loop analytic coupling function
\begin{equation}
\bar{a}^{(1)}_{\an}(Q^2/\Lambda^2)=\frac{1}{\beta_0}\left[\frac{1}
{\log(Q^2/\Lambda^2)}+\frac{\Lambda^2}{\Lambda^2-Q^2}\right].
\label{2.4}
\end{equation}
The first term on the right-hand side preserves the standard
UV-behavior of the invariant coupling.  The second term, which comes
from the spectral representation and enforces the proper analytic
properties, compensates the ghost pole at $Q^2=\Lambda^2$ and is
essentially nonperturbative (see the general discussion of this point
in~\cite{19}).  This term gives no contribution to the Taylor series
expansion.  Thus, the causality and spectrality principles expressed
in the form of $Q^2$-analyticity, send us the message that
perturbation theory is not the whole story.  The requirement of proper
analytic properties leads to the appearance of contributions given by
powers of~$Q^2$ that cannot be seen in the original perturbative
expansion.  We note also that unlike in electrodynamics, the
asymptotic freedom property in QCD has the effect that such
nonperturbative contributions show up in the effective coupling
function already in the domain of low energies and momentum transfers
reachable in realistic experiments, rather than at unrealistically
high energies.

           \begin{figure}[hbt]
\centerline{ \epsfig{file=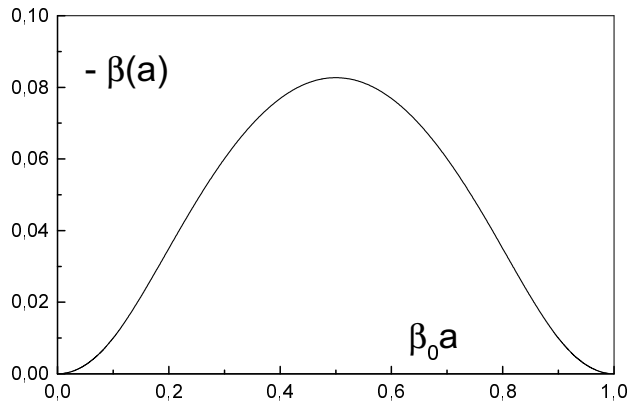,width=9.7cm}} \caption{{\sl The graph of
the one-loop $\beta$-function. }}
         \label{shir2}
         \end{figure}

Thus, synthesis of the renormalization-group invariance and analyticity
leads to the analytic invariant charge without the logarithmic pole and with
a finite IR value~\footnote{For numerical estimates at small~$Q^2$, we use
the number of active quarks~$f=3$.}
$\bar{\alpha}_{\an}(0)=4\pi/\beta_0\simeq1.396$.  This limiting value is
independent of the experimental information related to the normalization
point $a=a(\mu^2)$ or to the parameter~$\Lambda$; it is instead determined
only by the $\beta$-function coefficient related to the general group
structure of the Lagrangian.  Figure~1 shows a bundle of curves
$\bar{\alpha}_{\an}(Q^2)$ corresponding to different values of~$\Lambda$ and
also the standard solutions corresponding to the same~$\Lambda$.

The graph of the one-loop $\beta$-function illustrating the existence of an
infrared fixed point in the analytic approach is shown in Fig.\,2.  The
horizontal axis is the parameter $\beta_0a$ and the vertical axis is the
function $-\beta(a)$.  We note that in the one-loop case, one has the
symmetry with respect to the point $\beta_0a=1/2$, which is broken when
taking higher orders into account.

We now proceed to the two-loop case.  The corresponding $\beta$-function
reads as
\begin{equation}
\beta(a)=-\beta_0a^2(1+b_1a),\qquad
b_1=\frac{\beta_1}{\beta_0},\qquad
\beta_1=102-\frac{38f}{3}\, .
\label{2.5}
\end{equation}
Integrating the renormalization group equation, we obtain the
transcendental relation
\begin{equation}
\beta_0\log x=\frac{1}{\bar{a}(x)}-
b_1\log\left(1+\frac{1}{b_1\bar{a}(x)}\right)
\label{2.6}
\end{equation}
that can be solved in terms of the Lambert function~\cite{20,21}.

           \begin{figure}[hbt]
\centerline{ \epsfig{file=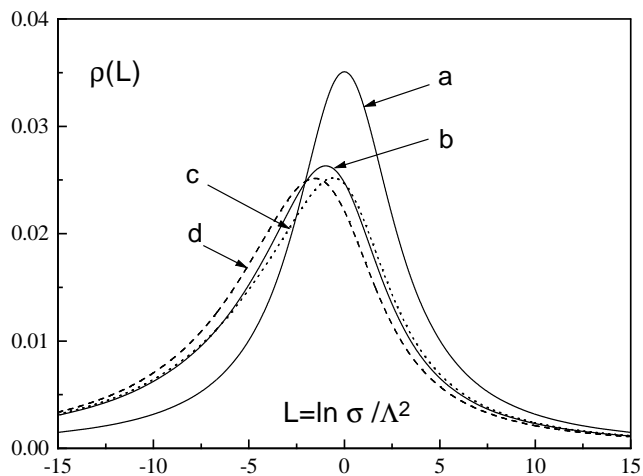,width=7.7cm}}
\caption{   {\sl
The spectral densities: the exact one-loop~({\sl
a\/}); the exact two-loop~({\sl b\/}); the approximate two-loop
expression (the iterative solution)~({\sl c\/}); the exact three-loop
function~({\sl d\/}). }}
        \label{shir3}
        \end{figure}

The spectral density obtained from this expression is shown in Fig.\,3
(curve~{\sl b}\/).  It proves to be very close to the spectral density
corresponding to the explicit iterative solution of Eq.~(\ref{2.6}),
\begin{equation}
\bar{a}^{(2)}(Q^2)=\frac{1}{\beta_0\ell+b_1\log(1+\beta_0\ell/b_1)},
\qquad\ell=\log\frac{Q^2}{\Lambda^2},
\label{2.7}
\end{equation}
which is useful in the subsequent analysis.

Solution~(\ref{2.7}) corresponds to the spectral function
\begin{equation}
\label{2.8}
\beta_0 \rho^{(2)}(\sigma)\,=\,\frac{I(L)}{R^2(L)\,+\,I^2(L)}\, ,
\quad L=\ln\frac{\sigma}{\Lambda^2}\, ;
\end{equation} %
\begin{eqnarray}
\label{2.9}
R(L)&=&L+B_1\ln \sqrt{\left(1+\frac{L}{B_1}\right)^2+
\left(\frac{\pi}{B_1} \right)^2}~,
\\ \nonumber
I(L)&=&\pi+B_1{\rm arccos}\frac{B_1+L}
{\sqrt{\left(B_1+L\right)^2+\pi^2}}~,\qquad B_1=\frac{\beta_1}{\beta_0^2}\, .
\end{eqnarray}
Its graph is given in Fig.\,3 (curve~{\sl c}\/), where we also show
the one-loop (curve~{\sl a}\/) and the three-loop (curve~{\sl d}\/)
results.  The three-loop~$\rho^{(3)}$ shown in Fig.\,3 is obtained in
the $\overline{\rm MS}$-scheme from the exact integral of the RG-equation
with the three-loop coefficient
$$
\beta_2\,=\,\frac{2857}{2}-\frac{5033}{18}\,f+\frac{325}{54}\,f^2\,
\,\,\stackrel{f=3}{=}\,\,
\,\frac{3863}{6}\simeq 643.83 \, . $$

As can be seen from Fig.\,3, the behavior of spectral densities is
stabilized starting with the two-loop level; as shown in what follows,
moreover, the areas below each of these curves are the same, which
corresponds to the universality of~$\bar{a}_{\an}(0)$.

           \begin{figure}[htb]
\centerline{ \epsfig{file=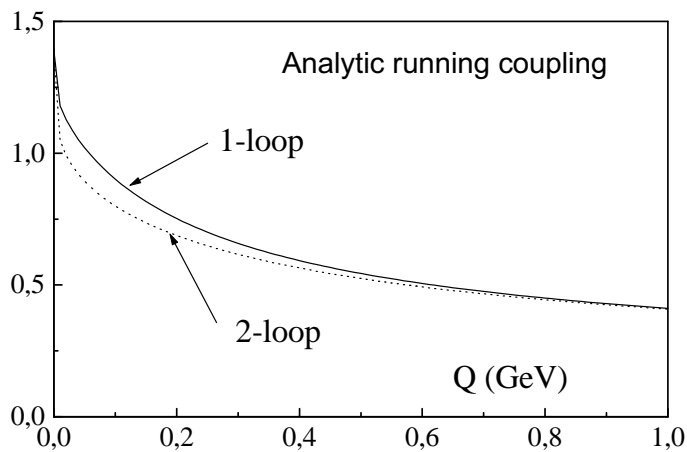,width=10.7cm}} \caption{ {\sl
Stability of the analytic invariant charge with respect to higher-loop
corrections.  We use the normalization at the $\tau$-lepton mass
$\bar{\alpha}_{\an}(M_{\tau}^2)=0.34$ for $f=3$; 1~is the one-loop
approximation, 2~is the two- and three-loop approximations. }}
        \label{shir4}
    \end{figure}

To obtain $\bar{a}^{(2)}_{\an}(Q^2)$, we have to insert spectral
density~(\ref{2.8}) in Eq.~(\ref{2.1}).  The resulting integral cannot be
evaluated explicitly.\footnote{In what follows, we explicitly give the
corresponding approximate formulas.}  The proper analytic properties are
reconstructed by not only eliminating the pole, but also by subtracting the
unphysical branch cut $0<Q^2<\Lambda^2\exp(-B_1)$ caused by the
double-logarithm dependence in~(\ref{2.7}).

The numerical calculation results for $f=3$ and for the normalization at the
point $\bar{\alpha}_{\an}(M_{\tau}^2)=0.34$ are shown in~Fig.\,4, where we
also give the one-loop curve (the corresponding values of~$\Lambda$ are
given in Table~1).  The three-loop $\overline{\rm MS}$-curve is practically
identical with the two-loop one, with the accuracy of the order~$1\,\%$.
Thus, in contrast with perturbation theory, analyticity leads to an
essential stabilization of the invariant charge behavior in the IR region.
Recalling the asymptotic freedom property, we obtain stability in all the
Euclidean domain~$0<Q^2<\infty$.

We note here that the universal behavior of the analytic coupling
function is not a consequence of the particular two-loop
formula~(\ref{2.7}).  The same conclusion remains valid when using
the exact solution~(\ref{2.6}).  Thus, the IR stability of the
analytic charge is an internal property of the method and is ensured
by contributions that are not analytic in~$\alpha_s$.  This approach
does not introduce any additional parameters into the theory; it
operates only with the scaling parameter~$\Lambda$ or with a certain
normalization point.

\subsection{Subtraction of unphysical singularities}

The analytic expression for the invariant coupling was obtained using
spectral representation~(\ref{2.1}) that guarantees the proper
analytic properties in the complex $Q^2$ plane and effectively amounts
to subtracting the unphysical singularities (the pole and the cuts).
It it useful to explicitly separate these terms.

We consider the complex plane of $z=Q^2/\Lambda^2$.  The method of
subtracting the singularities allows us to obtain an explicit
expression for the analytic coupling in the one-loop case.  Indeed,
the expression $\beta_0\bar{a}^{(1)}(z)={1}/{\log z}$ has an
unphysical pole at $z=1$ with the residue
$\rres\,\bigl[\beta_0\bar{a}^{(1)}(z),z=1\bigr]=1$, whose elimination
amounts to adding the term $1/(1-z)$, such that the expression
satisfying the proper analytic properties has the form given
in~(\ref{2.4}).

           \begin{figure}[hbt]
\centerline{ \epsfig{file=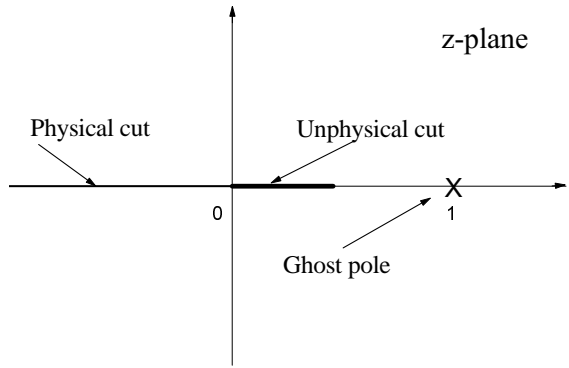,width=9.7cm}}
\caption{{\sl Two-loop
singularities in the complex plane ($z=Q^2/\Lambda^2$). }}
        \label{shir5}
        \end{figure}

In the two-loop case, we first consider~(\ref{2.7}), which in
addition to having the ghost pole at $z=1$ with the residue
$\rres \, \bigl[\beta_0\bar{a}^{(2)}(z),z=1\bigr]=1/2$, has an unphysical
cut along the positive part of the real axis $0<z<\exp(-B_1)$
(see~Fig.\,5).  The subtraction is effected by the pole term
\begin{equation}
\beta_0\Delta\bar{a}^{(2)}_{\pole}(z)=\frac{1}{2}\frac{1}{1-z}
\label{2.10}
\end{equation}
and by the integral

\begin{eqnarray}
\label{2.11}
\beta_0\Delta\bar{a}^{(2)}_{\cut}(z)&=&
\frac{1}{\pi}\int_0^{\exp(-B_1)}\frac{d\sigma}{\sigma-z}\\
&\times& \frac{\pi
B_1}{\bigl[\log(\sigma)+
B_1\log\bigl(-1-\log(\sigma)/B_1\bigr)\bigr]^2+\pi^2B_1^2}
\nonumber
\end{eqnarray}
that eliminates the unphysical branch cut.  As the result, the
analytic invariant charge can be written as
\begin{equation}
\beta_0\bar{a}_{\an}^{(2)}(z)=\beta_0\bar{a}^{(2)}(z)+
\beta_0\Delta\bar{a}^{(2)}_{\pole}(z)+
\beta_0\Delta\bar{a}^{(2)}_{\cut}(z).
\label{2.12}
\end{equation}
This form is convenient because the analytic coupling is
represented as a sum of the standard expression and of the additional
terms of a nonperturbative nature.  Their contribution can be
represented as an expansion in powers of~$\Lambda^2/Q^2$ [see
Eq.~(\ref{2.22}) below].

\subsection{Universality of $\bar a_{\an}(0)$ }
The universal value at $Q^2=0$ is formed by the contribution of the pole
term $\Delta\bar{a}^{(2)}_{\pole}(z=0)=1/(2\beta_0)$ and the
contribution of~(\ref{2.11}) that can be represented as
$$
\Delta\bar{a}^{(2)}_{\cut}(z=0)=\frac{1}{\beta_0}
\int_0^{\infty}\frac{dx}{(x+1-\log x)^2+\pi^2}=\frac{1}{2\beta_0}\, .
$$
The total contribution leads to the universal expression
$\bar{a}_{\an}(0)=1/\beta_0$.

In the above approach to approximating the original two-loop coupling,
the residue at the pole (which is the leading unphysical singularity)
is independent of the two-loop $\beta$-function coefficient, and it
may thus seem that precisely this fact makes $\bar{a}_{\an}(0)$
independent of higher-loop corrections.  As we have noted, however,
there is a different reason behind the universality of
$\bar{a}_{\an}(0)$, which does not reduce to the choice of a
particular approximation of the original invariant coupling.  We now
explain this in more detail.  The standard asymptotic two-loop
expression can be obtained by expanding the function
\begin{equation}
\beta_0\bar{a}^{(2)}(z)=
\frac{1}{\log z+B_1\log\bigl[1+\log(z/C)\bigr]}\, ,
\label{2.13}
\end{equation}
where~$C$ is a constant.  Expression~(\ref{2.13}) correctly
reproduces the standard UV limit
\begin{equation}
\bar{\alpha}_s=\frac{4\pi}{\beta_0}\,
\left[\frac{1}{\log(Q^2/\Lambda^2)}\,-\,
\frac{\beta_1}{\beta_0^2}\,\frac{\log\log(Q^2/\Lambda^2)}
{\log^2(Q^2/\Lambda^2)}\right]
\label{2.14}
\end{equation}
that is independent of the constant~$C$. At the same time, the
residue at the pole now depends on the two-loop $\beta$-function
coefficient through~$B_1$,
\begin{equation}
\rres \,\bigl[\beta_0\bar{a}^{(1)}(z),z=1\bigr]=\frac{1}{1+B_1/C}\, ,
\label{2.15}
\end{equation}
and therefore, the same dependence is involved in the corresponding
compensating term
\begin{equation}
\beta_0\Delta\bar{a}^{(2)}_{\pole}(z)=\frac{1}{1+B_1/C}\,\frac{1}{1-z}
\label{2.16}
\end{equation}
whose contribution to $\bar{a}_{\an}(0)$ is equal to
\begin{equation}
\Delta\bar{a}^{(2)}_{\pole}(0)=\frac{1}{\beta_0}\,\frac{1}{1+B_1/C}.
\label{2.17}
\end{equation}

The contribution to $\bar{a}_{\an}(0)$ of the term compensating the
unphysical branch cut is now given by the integral
\begin{eqnarray}
\label{2.18}
\Delta\bar{a}^{(2)}_{\cut}(z=0)&=&\frac{1}{\beta_0}\,\frac{C}{B_1}\,
\int_0^{\infty}\frac{dx}{\bigl[(x+1)C/B_1-\log x\bigr]^2+\pi^2}
\nonumber\\
&=&\frac{1}{\beta_0}\,\frac{B_1}{(B_1+C)},
\end{eqnarray}
which together with the pole contribution~(\ref{2.17}) gives the
universal value $\bar{a}_{\an}(0)=1/\beta_0$ that is independent of
either~$C$ or~$B_1$.

When taking the higher-loop contribution into account for proving the
universality of the IR limit in the analytic approach, it is
convenient to use the complex quantity $\zeta=1/a$.  We now give
simpler arguments based on the expansion of the perturbative charge
into a double series in powers of $\log^m(\ell)/\ell^k$.  For
$\bar{a}_{\an}(0)$, we can write
\begin{equation}
\bar{a}_{\an}(0)=\frac{1}{\pi}\int_{-\infty}^{\infty}dL\,\varrho(L)=
\frac{1}{\beta_0}\,+\,\sum_{k=1}^{\infty}\sum_{m=0}^k
\alpha_{k,m}\,\Delta\bar{a}_{k,m}(0)\,,
\label{2.19}
\end{equation}
where the higher-loop contribution is given by
\begin{equation}
\Delta\bar{a}_{k,m}(0)\,=\,\frac{1}{\pi}\,\Ima
\int_{-\infty}^{\infty}dL\frac{\log^m(L-i\pi)}{(L-i\pi)^{k+1}}\,.
\label{2.20}
\end{equation}
Since the integrand in~(\ref{2.20}) has no singularities in the
lower half-plane, we immediately obtain $\Delta\bar{a}_{k,m}(0)=0$,
which proves the universality of the infrared fixed point value of the
analytic charge.

Thus, the analyticity requirement for the running charge leads to
essential modifications of perturbation theory in the IR region.  The
most relevant factor here is the universality of the IR limiting value
of the analytic coupling function (the invariance with respect to
higher-loop corrections), which results in that the family of the
invariant charge curves corresponding to different loop approximations
looks as a bundle with the common point at $Q^2=0$.  In addition, these
curves obviously come closer to each other in the UV region in view of
the asymptotic freedom property.  In our approach, unlike in the
standard perturbation theory, there emerges a remarkably stable
picture of the invariant charge behavior with respect to higher
corrections.  This stability is important for phenomenological
applications, where the relevant energy interval is of the order of or
less than several~GeV.

\subsection{Approximate formulas}

The explicit one-loop formula~(\ref{2.4}) is very simple, and its use does
not lead to any complications. In the two-loop case, the analytic coupling
is written in the form of an integral representation, and it is interesting
to find explicit approximate expressions that are convenient in
applications.

We consider two such formulas.  The first expression follows directly
from the picture of subtracting the unphysical singularities as
explained in Sec.~3.  Thus, the analytic coupling can be represented
as
\begin{equation}
\bar{\alpha}_{\an}(Q^2)=\bar{\alpha}_{\TV}(Q^2)+
\Delta\bar{\alpha}_{\rm sing}(Q^2)\, ,
\label{2.21}
\end{equation}
where $\bar{\alpha}_{\TV}(Q^2)$ is a perturbative contribution and
the term $\Delta\bar{\alpha}_{\sing}(Q^2)$ has the effect of
subtracting the unphysical singularities.  For the perturbative term
taken as in~(\ref{2.7}), the term eliminating the unphysical
singularities can be represented as two terms whose respective effects
are to subtract the unphysical pole and the branch cut.  The term
compensating the pole has a simple form.  For the term compensating
the cut, we use the fact that the expansion coefficients~$C_k$,
\begin{eqnarray}
\label{2.22}
\Delta\bar{\alpha}_{\cut}(Q^2)&=&-\frac{4\pi}{\beta_0}
\sum_{k=1}^{\infty}{\left(\frac{\Lambda^2}{Q^2}\right)}^kC_k\,,\\[0.25cm]
C_k&=&\int_0^{\infty}dt\,
\frac{\exp\bigl[-B_1k(t+1)\bigr]}{(t+1-\log t)^2+\pi^2}\, ,\nonumber
\end{eqnarray}
are numerically small and decrease rapidly ($C_1=0.0354$,
$C_2=0.0079$, $C_3=0.0023$, \dots).  Keeping only the first term in
the expansion, we obtain a simple interpolation formula
\begin{eqnarray}
\label{2.23}
\bar{\alpha}_{\appr}^{(2)}(Q^2)&=&\frac{4\pi}{\beta_0}
\biggl\{\frac{1}{\log(Q^2/\Lambda^2)+
B_1\log\bigl[1+\log(Q^2/\Lambda^2)/B_1\bigr]} \nonumber\\[0.25cm]
&&~~~\,-\,
\frac{1}{2}\frac{\Lambda^2}{Q^2-\Lambda^2}-
\frac{\Lambda^2}{Q^2}C_1\biggr\}\, ,
\end{eqnarray}
which provides good approximation~\footnote{The approximate formula
for the two-loop correction $[\bar{a}^2]_{\an}$ to the physical
quantities of the $D$-function type can also be found in this way.} to
the two-loop analytic coupling for moderately large~$Q^2$.  In the
interval $1<Q<1.5\,$GeV, the accuracy of the approximation is not
worse than $0.4\,\%$, and for large values of~$Q$, the difference
between the formulas becomes negligible.  Thus,
expression~(\ref{2.23}) is quite acceptable in the domain of
moderately large $Q\geq1\,$GeV.

In a number of cases, however, it is necessary to deal with smaller
values of~$Q$, down to~$Q\simeq0$.  Formula~(\ref{2.23}) is no longer
applicable to such problems because the term compensating the branch
cut is poorly approximated by power-series expansion~(\ref{2.22}).
The approximate formula
\begin{eqnarray}
\label{2.24}
\bar{\alpha}_{\appr}^{(2)}(Q^2) &=& \frac{4\pi}{\beta_0}\left[
\frac{1}{\ell_2(Q^2)}+\frac{1}{1-\exp\bigl[\ell_2(Q^2)\bigr]}\right],
\\[0.25cm]
\ell_2(Q^2)\,&=&\,\ln\frac{Q^2}{\Lambda^2}\,+\,B_1\,
\ln\sqrt{\ln^2\frac{Q^2}{\Lambda^2}\,+\,4\pi^2}  \nonumber
\end{eqnarray}
for the two-loop analytic charge can be used also for $Q\simeq0$.
Equation~(\ref{2.24}) reproduces the UV two-loop asymptotic
behavior~(\ref{2.14}) and the universal limiting value at $Q^2=0$.
This expression approximates the exact one for $Q\geq1\,$GeV with the
accuracy within $1\,\%$ and can be used for all~$Q^2$.

\begin{table}
\caption{\sl The perturbative and analytic one- and two-loop
values of the scaling parameter (MeV) for $f=3$ versus the
normalization point $\bar{\alpha}_s(M_{\tau}^2)$.}

\hphantom{}

\begin{center}
\begin{tabular}{|c|c|c|c|c|c|}  \hline
$~~~\bar{\alpha}_s(M_{\tau}^2)~~~$
&0.30  &0.32  &0.34  &0.36  &0.38  \\ [0.1cm] \hline
$~~\Lambda_{\rm PT}^{(1)}~~$
&~~~173~~~&~~~201~~~&228~~~&~~~256~~~&~~~283~~~     \\ [0.1cm] \hline
$\Lambda_{\rm an}^{(1)}$      &197&235&275&319&366     \\ [0.1cm]
\hline
$\Lambda_{\rm PT}^{(2)}$      &333&377&419&460&500     \\ [0.1cm]
\hline
$\Lambda_{\rm an}^{(2)}$      &434&516&607&706&814     \\ [0.1cm]
\hline
$\Lambda_{\rm appr}^{(2)}$   &423&500&582&671&777      \\
\hline
\end{tabular}
\end{center}
\end{table}

For sufficiently large~$Q^2$, the analytic coupling function is
dominated by its perturbative component.  Already for $Q=M_{\tau}$,
however, the nonperturbative contribution becomes essential.  In
Table~1, we compare the~$\Lambda$ parameter values corresponding to
the perturbative and analytic approaches.  The result obtained
according to Eq.~(\ref{2.23}) reproduces the exact two-loop
calculation with high accuracy and is not given given here.  For the
two-loop perturbative formula, we used expression~(\ref{2.7}), which
is most appropriate for our analysis.\footnote{We note that using
formula~(\ref{2.7}) as a perturbative one, leads to somewhat greater
values of~$\Lambda$ than when working it out from Eq.~(\ref{2.14}).}
The bottom row corresponds to approximate expression~(\ref{2.24}).

\section{Analytic perturbation theory} \label{sec3}

In this section, we briefly review applications of the analytic
approach to the analysis of several processes.  For the physical
quantities considered here, we use the analyticization procedure of
the entire perturbative expression involving higher powers of the
invariant charge~\cite{9}.  This strategy leads to the so-called {\sl
analytic perturbation theory} (APT).

We consider the integral characteristics of the invariant charge in
the IR region by extracting the relevant information from the physics
of jets, and also from the $e^+e^-$-annihilation processes into
hadrons and the inclusive $\tau$-lepton decay.  We use this set of
data to study the dependence of theoretical results on the choice of
the renormalization scheme.  We show that applying the APT allows us
to considerably reduce the scheme dependence.  This in turn means that
the three-loop level attained for many processes is practically
independent of the choice of the scheme.

\subsection{The integral characteristics of $\bar\alpha_s$ in the IR
region}

A distinctive feature of the analytic charge is that it is finite in the IR
region.  This property, which is sometimes referred to as the coupling
``freezing,'' is often used for phenomenological purposes (see, for example,
the discussion in~\cite{22}). Experimental evidence for the regular IR
behavior of the QCD charge was ingeniously extracted from physics of jets
using the integral characteristics
\begin{equation}
A(Q)=\frac{1}{Q}\int_0^Qdk\,{\bar{\alpha}_s}(k^2).
\label{3.1}
\end{equation}
It has been empirically found~\cite{23} that
$A(2\,{\mbox{\rm GeV}})=0.52\pm0.10$.

\begin{table}[bh]
\caption{\sl The infrared integral characteristics of
$\bar{\alpha}_s(k^2)$ evaluated in the one- and two-loop
approximations for normalization at the $\tau$-lepton mass.}
\hphantom{}
\begin{center}
\begin{tabular}{|c|c|c|c|}  \hline
~$\bar{\alpha}_{\rm an}(M_{\tau}^2)~$ & 0.34  &0.36 &0.38 \\  [0.1cm]
\hline
~~~$A_{\rm 1-loop}(2\,{\rm Ē'})$~~~&~~~0.50~~~ &~~~0.52~~~&~~~
0.55~~~ \\ [0.1cm]
 \hline
~~~$A_{\rm 2-loop}(2\,{\rm Ē'})$~~~&~~~0.48~~~ &~~~0.50~~~&~~~0.52~~~
\\  \hline
\end{tabular}
\end{center}
\end{table}

We normalize $\bar{\alpha}_s$ at the $\tau$-lepton mass. Calculations of
$A$(2\,GeV) are given in Table~2.  It can be seen that the APT approach
allows us to uniformly and consistently describe the almost-perturbative
region of the order of the $\tau$-lepton mass and the nonperturbative
characteristics~(\ref{3.1}) without introducing any additional parameters.

\subsection{The $e^+e^-$-annihilation process into hadrons}

We now apply the analytic approach to the analysis of the
$e^+e^-$-annihilation process into hadrons.  To compare the results with the
experimental data, we use the method of so called ``smearing'' of resonances
proposed in~\cite{24}.  The analysis of the $e^+e^-$-annihilation into
hadrons carried out in~\cite{22} relied on a certain ``optimum''
renormalization scheme constructed on the base of the principle of minimal
sensitivity (PMS)~\cite{26} with the third-order perturbation theory used
for optimization.  Our analysis is not based on any optimization of the
scheme arbitrariness.  Moreover, we show that the scheme dependence in the
APT is considerably less than in the standard approach, and its predictions
have practically no scheme arbitrariness in the entire energy range.

The analyticization procedure can be also applied to observable
quantities for which the appropriate analytic properties are known.
The APT can be applied to an object that has numerous applications,
namely the Adler $D$-function
\begin{equation}
D(Q^2)=-Q^2\frac{d\Pi(-Q^2)}{dQ^2}=3\sum_{f}Q_f^2\bigl[1+d(Q^2)\bigr],
\label{3.2}
\end{equation}
where $\Pi(s)$ is the correlation function and $d(Q^2)$ is the
QCD correction that is expanded in the RG perturbation theory as
\begin{equation}
d(Q^2)=a(Q^2)\bigl[1+d_1a(Q^2)+d_2a^2(Q^2)+\cdots\bigr],
\label{3.3}
\end{equation}
where~\footnote{In this formula, we allowed ourselves to change the
normalization of the coupling constant so as to simplify comparing
with the previous works on the subject, where, as a rule, the quantity
$a=\alpha_s/\pi$ is used as the invariant charge.}
$a=\alpha_s/\pi$.

The $D$-function is related to the function $R(s)$ defined as the
ratio of the hadron and lepton cross-sections for the
$e^+e^-$-annihilation by
\begin{equation}
D(Q^2)=Q^2\int_0^{\infty}\frac{ds}{{(s+Q^2)}^2}R(s).
\label{3.4}
\end{equation}
This also implies the properties of $D(Q^2)$ as an analytic function in the
$Q^2$-plane cut along the negative semi-axis.  We define the spectral
density $\rho^{\eff}(\sigma)$ through the discontinuity of~(\ref{3.3}) on
this cut,
\begin{equation}
\rho^{\eff}(\sigma)=\rho^{(1)}(\sigma)+d_1\rho^{(2)}(\sigma)+
d_2\rho^{(3)}(\sigma)+\cdots.
\label{3.5}
\end{equation}
The expression $\rho^{(1)}(\sigma)$ is the spectral function of the
invariant charge and $\rho^{(k)}(\sigma)$ in~(\ref{3.5}) corresponds
to the $k$th power of the effective coupling.  Thus, the analytic
expression for the QCD correction to the $D$-function is written as
\begin{equation}
d_{\ATV}(Q^2)={\delta_{\ATV}^{(1)}(Q^2)}+d_1{\delta_{\ATV}^{(2)}(Q^2)}+
d_2{\delta_{\ATV}^{(3)}(Q^2)}+\cdots,
\label{3.6}
\end{equation}
where the first term $\delta_{\ATV}^{(1)}(Q^2)$ coincides with the analytic
invariant charge.  The subsequent terms do not reduce to powers of the
analytic coupling; thus, the APT method leads to non-power-series
expansions.  Properties of such expansions were analyzed in~\cite{12}.

We define the QCD correction $r(s)$ to the function $R(s)$ in the same
manner as for the $D$-function in~(\ref{3.2}), and we use the
relations
\begin{equation}
d(Q^2)=Q^2\int_0^{\infty}\frac{ds}{{(s+q^2)}^2}r(s),\qquad
r(s)=-\frac{1}{2\pi i}\int_{s-i\epsilon}^{s+i\epsilon}\frac{dz}{z}d(-z)\,,
\label{3.7}
\end{equation}
where the integration contour in the last expression is in the
analyticity domain of the integrand and bypasses the cut along the
real semi-axis.

We take the quark thresholds into account by using the approximate
formula proposed in~\cite{24},
\begin{equation}
R(s)=3\sum_{f}\, Q_f^2\, \theta(s-4m_f^2)\,T(v_f)\,
\bigl[1\,+\,g(v_f)\, r_f(s)\bigr]\, ,
\label{3.8}
\end{equation}
where $v_f$ and the functions $T(v)$ and $g(v)$ are given by

\begin{eqnarray}
\label{3.9}
v_f&=&\sqrt{1-\frac{4m_f^2}{s}}\,,\qquad
T(v)=\frac{v(3-v^2)}{2}\, , \nonumber \\[0.3cm]
g(v)&=&\frac{4\pi}{3}\left[\frac{\pi}{2v}-\frac{3+v}{4}
\left(\frac{\pi}{2}-\frac{3}{4\pi}\right)\right]\, .
\end{eqnarray}

In the APT, the correction $r_f(s)$ is expressed through the effective
spectral density as
\begin{equation}
r_f(s)\,=\,\frac{1}{\pi}\,\int_s^{\infty}\frac{d\sigma}{\sigma}\,
\rho_f^{\eff}(\sigma),
\label{3.10}
\end{equation}
where $\rho_f^{\eff}(\sigma)$ is defined in terms of the discontinuity of
$d_f(Q^2)$ on the physical branch cut. The corresponding three-loop
contribution is written as
\begin{equation}
d_f(Q^2)=a_f(Q^2)\bigl[1+d_f^{(1)}a_f(Q^2)+d_f^{(2)}a_f^2(Q^2)\bigr],
\label{3.11}
\end{equation}
where the $\overline{\rm MS}$-scheme coefficients are equal to~\cite{26}
\begin{eqnarray}
\label{d_f(i)}
d_f^{(1)}&=&1.986-0.115\, f\, ,\nonumber\\
d_f^{(2)}&=&18.244-4.216\,
f+0.086\,f^2+d_f^{\rm singlet} \, , \nonumber \\
d_f^{\rm singlet}&=& -\frac{1.2395}{3}\,
\frac{(\sum_{f'}^fQ_{f'})^2}{\sum_{f'}^fQ_{f'}^2} \, .
\nonumber
\end{eqnarray}

It is hardly possible to use perturbative expressions for a direct
description of the experimentally observed quantity $R(s)$, because of the
threshold singularities of the form $(\alpha_s/v)^n$.  We use the
``smearing'' method proposed in~\cite{24}, which does nevertheless allow us
to compare the results with the experiment.  The idea of this approach
consists in replacing the quantity $R(s)$ defined through the correlation
function~$\Pi$ as
\begin{equation}
R(s)=\frac{1}{2i}\,\bigl[\Pi(s+i\epsilon)-\Pi(s-i\epsilon)\bigr],
\label{3.12}
\end{equation}
with the quantity
\begin{equation}
R_{\Delta}(s)=\frac{1}{2i}\,\bigl[\Pi(s+i\Delta)-\Pi(s-i\Delta)\bigr]
\label{3.13}
\end{equation}
for some finite~$\Delta$.  For the values of~$s$ near the
threshold, quantity~(\ref{3.12}) is very sensitive to the threshold
singularities, in the vicinity of which the perturbative expansion
looses its applicability.  Stepping away from the real axis into
the~$q^2$ complex plane by a finite distance~$\Delta$, as
in~(\ref{3.13}), we can expect that it would be possible to
describe~(\ref{3.13}) using an appropriate perturbative
approximation.

The ``experimental'' curve corresponding to~(\ref{3.13}) can be found
if we use the dispersion relation for the correlator $\Pi(q^2)$ to
write Eq.~(\ref{3.13}) as
\begin{equation}
R_{\Delta}(s)=\frac{\Delta}{\pi}\int_0^{\infty}ds'
\frac{R(s')}{(s-s')^2+\Delta^2}.
\label{3.14}
\end{equation}
The corresponding ``experimental'' curves were found in~\cite{22}
for some values of~$\Delta$, whose estimates were made in~\cite{24}.
We use these curves for comparing with our results.

We note that a direct use of perturbation theory for describing
$R_{\Delta}(s)$ is again impossible.  Indeed, the $R$-ratio
in~(\ref{3.14}) parametrized using the invariant charge with
unphysical singularities leads to a divergence of the integral
in~(\ref{3.14}).  Thus, even though the use of the ``smeared''
quantity~(\ref{3.14}) allows us to bypass the complication with the
threshold singularities, there arises a problem related to the
behavior of the running charge in the IR region.  We can avoid this
complication using the APT.

            \begin{figure}[htpb]
\centerline{ \epsfig{file=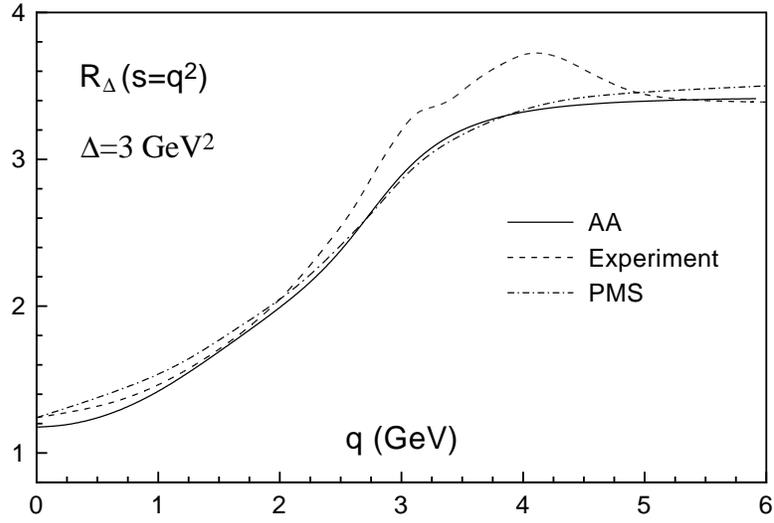,width=12.7cm}}
     \caption{    {\sl
The quantity $R_{\Delta}(q^2)$ corresponding to
the parameter value $\Delta=3\,{\mbox{\rm GeV}}^2$.  The figure shows the
experimental curve, the result of the PMS-optimization of the
third-order perturbative expansion obtained in~\protect\cite{22}, and the
result of the analytic approach through the third order. }}
\label{shir6}
          \end{figure}

For $\Delta=3\,{\mbox{\rm GeV}}^2$, Fig.~6 shows the corresponding
experimental curve and the curve found in~\cite{22} from the
PMS-optimization of the third-order perturbative expansion.  The same figure
gives also the result of our calculation through the third
order.\footnote{As shown in~\cite{8}, the calculation of $R_{\Delta}$ in the
analytic approach leads to good fit of the experimental curve already in the
first order.}  For the scaling parameter in the analytic approach, we took
the value $\Lambda_{\an}=870\,$MeV \ ($f=3$) obtained from the analysis of
the semileptonic $\tau$-decay in the APT framework.  For the quark masses,
we took the values that are close to the constituent ones (cf.~\cite{27}),
$m_u=m_d=250\,$MeV, \ $m_s=400\,$MeV, \ $m_c=1.35\,$GeV, \ $m_b=4.75\,$GeV
and $m_t=174\,$GeV.

\begin{figure}[hpbt]
\centerline{\epsfig{file=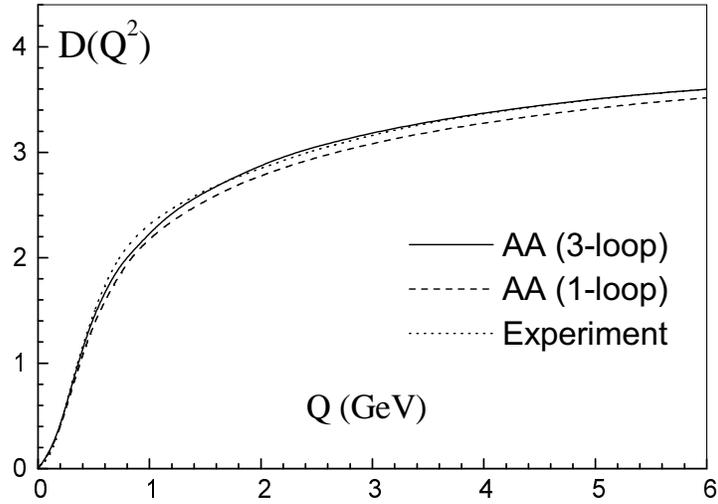,width=12.7cm}}
\caption{ {\sl $D$-function.  } }
\label{shir7}
\end{figure}

New ``experimental'' data for the $D$-function were obtained
recently~\cite{28}.  We give in Fig.\,7 the corresponding curves and
also the result of our calculation.  Figure~7 shows that good fit of
the experimental data can be achieved already in the first order of
APT.  The same conclusion, as we have noted, is valid for $R_{\Delta}$
in the entire energy range~\cite{8}. We note here that loop stability
is not observed in the standard approach using the PMS-optimization.
Moreover, the whole ``trick'' is based here on higher approximations.
Thus, the situation regarding the absence or the presence of the
infrared fixed point that can emerge in scheme optimization of the
perturbative expansion depends in an essential way on the quantity
under consideration (i.e., is defined by the coefficients of the
perturbative expansion)~\cite{29}.

\sloppy
\subsection{The dependence on the renormalization scheme}
\fussy

Inevitable termination of the PT series, i.e., the approximation of a
physical quantity by one of its partial sums, leads to the known problem of
the dependence of the results on the renormalization prescription.  Thus,
the partial sum of the PT series used in approximating a physical quantity
bears a dependence on the choice of the renormalization scheme, which is the
source of theoretical ambiguity in describing experimental data.  In QCD,
such ambiguity is the greater the smaller are the energy parameters
characteristic of the process.  To solve the stability problem of the
results obtained, it is by far not enough to investigate only loop stability
within a certain renormalization scheme; one should also consider the scheme
stability of the results.

We discuss the scheme arbitrariness arising in the APT in the example
of the $R$-ratio for the $e^+e^-$-annihilation process into hadrons.
We consider a class of MS-like schemes and compare our results with
those obtained in the perturbative analysis (see, for
example,~\cite{30}).

In passing from one renormalization scheme to another, the coupling
constant transforms as
\begin{equation}
a'=a(1+v_1a+v_2a^2+\cdots).
\label{3.15}
\end{equation}
We limit ourselves here to the three-loop level of the $D$-function
achieved at present, with the QCD corrections taken in the
approximation where
\begin{equation}
d=a(1+d_1a+d_2{a}^2),
\label{3.16}
\end{equation}
with the running charge determined as a solution of the
renormalization group equation with the three-loop $\beta$-function
\begin{equation}
\beta(a)=\mu^2\frac{\partial a}{\partial\mu^2}=
-ba^2(1+b_1a+b_2a^2)\, ,
\label{3.17}
\end{equation}
where
\begin{eqnarray}
\label{3.18}
b&=&\frac{33-2f}{6}\, ,\qquad b_1=\frac{153-19f}{66-4f}, \\[0.25cm]
b_2^{\overline{\rm MS}}&=&\frac{77139-15099f+325f^2}{288(33-2f)}\, .
\nonumber
\end{eqnarray}

The three-loop $\beta$-function coefficient~$b_2$ and the expansion
coefficients~$d_1$ and~$d_2$ in~(\ref{3.16}) depend on the choice of
the renormalization scheme.  Under scheme
transformation~(\ref{3.15}), they change as
\begin{eqnarray}
\label{3.19}
b_2'&=&b_2-v_1^2-b_1v_1+v_2\, ,               \nonumber \\
d_1'&=&d_1-v_1\, ,                            \\
d_2'&=&d_2-2(d_1-v_1)v_1-v_2\, .              \nonumber
\end{eqnarray}

Thus, every term in representation~(\ref{3.16}) undergoes a
transformation, and we thus obtain the new function
\begin{equation}
d'=a'(1+d_1'a'+d_2'{a'}^2),
\label{3.20}
\end{equation}
where the coupling $a'$ is evaluated with the new $\beta$-function,
with the three-loop coefficient~$b_2$ replaced by the primed
one~$b_2'$.

Recalling the transformation law of the scaling parameter~\cite{31}
$$\Lambda'=\Lambda\exp({v_1}/{b})$$ and Eqs.~(\ref{3.19}), we
find two scheme invariants \cite{25}
\begin{equation}
\rho_1=\frac{b}{2}\log\frac{Q^2}{\Lambda^2}-d_1,\qquad
\rho_2=b_2+d_2-b_1d_1-d_1^2.
\label{3.21}
\end{equation}

We normalize the momentum scale at $\Lambda_{\overline{\rm MS}}$.  In
arbitrary scheme, the invariant charge is then determined from the
equation
\begin{equation}
\frac{b}{2}\,\log\left(\frac{Q^2}{\Lambda_{\overline{\rm MS}}^2}\right)=
d_1^{\overline{\rm MS}}-d_1+\Phi(a,b_2)\, ,
\label{3.22}
\end{equation}
where
\begin{equation}
\Phi(a,b_2)=\frac{1}{a}-b_1\log\frac{1+b_1a}{\beta_0a}+
b_2\int_0^a\frac{dx}{(1+b_1x)(1+b_1x+b_2x^2)}.
\label{3.23}
\end{equation}

Although there are no general arguments to prefer a certain
renormalization scheme from the start, we nevertheless can define a
class of ``natural'' schemes, which look reasonable at the three-loop
level that we consider.  The relevant criterion was proposed
in~\cite{32}.  One should restrict oneself to the schemes where the
cancellations between different terms in the second scheme
invariant~(\ref{3.21}) are not too large.  Quantitatively, this
criterion can be related to the cancellation index
\begin{equation}
C=\frac{1}{|\rho_2|}\,\bigl(|b_2|+|d_2|+d_1^2+|d_1|b_1\bigr)\, .
\label{3.24}
\end{equation}
One should of course keep in mind the conventions involved in these
considerations, in particular as regards the minimal value of the
cancellation index.

Given a certain maximum value of the cancellation index $C_{\max}$, we
can investigate stability of the results obtained by taking different
schemes with the index $C\leq C_{\max}$.  As $C_{\max}$, we take the
index corresponding to the optimal PMS-scheme.  We then have a
relatively small class of ``admissible'' schemes bounded by the
maximal index~$C_{\PMS}$.

For $R(s)$, the cancellation index~$C_R$ is evaluated using the known
coefficients $r_1$ and~$r_2$ of the perturbative expansion of the
correction $r=a(1+r_1a+r_2a^2)$.  For the PMS-scheme, it
is~$C_{\PMS}\simeq2$.  To demonstrate the scheme arbitrariness arising
here, we choose two schemes from this class.  The first one is the~$H$
scheme with the parameters $r_1^{({\rm H})}=-3.2$ and $b_2^{({\rm H})}=0$
(the 't~Hooft scheme), and the second is the $\overline{\rm MS}$-scheme
corresponding to the parameters $r_1^{(\overline{\rm MS})}=1.64$ and
$b_2^{(\overline{\rm MS})}=4.47$.  These schemes are close to each other
and to the boundary cancellation index $C_{\rm H}\simeq
C_{\overline{\rm MS}}\simeq C_{\rm PMS}\simeq2$.

            \begin{figure}[hpt]
\centerline{ \epsfig{file=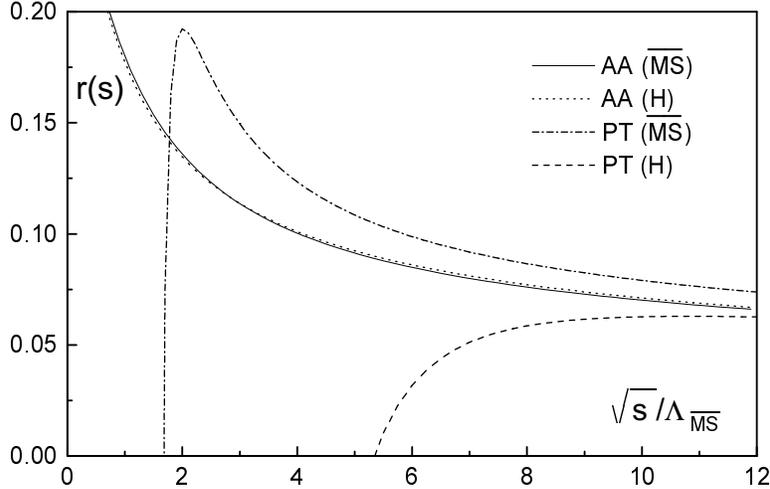,width=12.0cm}}
     \caption{  {\sl
The graph of $r(s)$ calculated in perturbation
theory (PT) and in the analytic approach (AA) for two renormalization
schemes~$H$ and $\overline{\rm MS}$ with the same cancellation
index~$C_R\simeq2$. }}
\label{shir8}
          \end{figure}

Figure~8 shows the QCD correction $r(s)$ as a function of
$\sqrt{s}/\Lambda_{\overline{\rm MS}}$ evaluated in perturbation theory and
in the analytic approach for two renormalization schemes~$H$ and
$\overline{\MS}$ with approximately the same cancellation indices
$C_R\simeq2$.  As can be seen from the figure, the analytic approach
allows us to drastically reduce the scheme arbitrariness.

Essential reduction of the scheme dependence in the APT also takes
place for other processes, for example the inclusive
$\tau$-decay~\cite{11}, and in the Bjorken and Gross--Llewellyn~Smith
sum rules for the inelastic lepton--hadron scattering~\cite{13,14}. In
the analytic approach, therefore, the three-loop level reached
presently for a number of physical processes is practically invariant
with respect to the choice of the renormalization prescription.

\subsection{Inclusive $\tau$-lepton decay}

The inclusive $\tau$-decay (see Fig.~9 for the corresponding diagram)
allows one to perform a low-energy test of~QCD.  The $\tau$-lepton
mass $M_{\tau}=1777^{+0.29}_{-0.26}\,$MeV \cite{33}, on the one hand,
is sufficiently large to allow the hadronic decay modes, but on the
other hand, is small in the chromodynamics scale, where it is in the
low-energy domain.  Theoretical description of the inclusive
$\tau$-decay is in principle possible without any model assumptions,
which is important for reliably determining the low-energy value
of~$\bar{\alpha}_{s}(M_{\tau}^2)$ from experimental data.  The main
quantity to be studied is the $R_{\tau}$-ratio
\begin{equation}
\label{3.25}
R_{\tau}= \frac{\Gamma\bigl[\tau^-\to\nu_{\tau}+{\rm
hadrons}(\gamma)\bigr]}
{\Gamma\bigl[\tau^-\to\nu_{\tau}e^-\overline{\nu}_{e}(\gamma)\bigr]},
\end{equation}
which in the modern experiments can be measures with the accuracy
of several per cent.

           \begin{figure}[htb]
~\hspace{-4.5cm}
\centerline{\epsfig{file=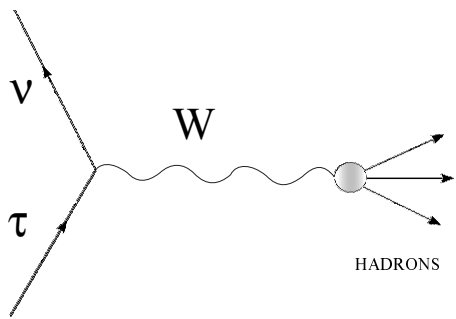,width=5.5cm,angle=90}}
\caption{ {\sl
The inclusive~$\tau$-lepton decay diagram.
 }                        }
        \label{shir9}
        \end{figure}

The starting point of the theoretical analysis is the expression
\begin{equation}
R_{\tau}=2\int_{0}^{M^{2}_{\tau}}\frac{ds}{M^{2}_{\tau}}
\left(1-\frac{s}{M^{2}_{\tau}}\right)^{2}
\left(1+\frac{2s}{M^{2}_{\tau}}\right)
\widetilde{R}(s),
\label{3.26}
\end{equation}
where $\widetilde{R}(s)$ is defined by the imaginary part of the
hadron correlator
\begin{equation}
\Pi(s)=\sum_{q=d,s}{\vert V_{uq}\vert}^2
\bigl[\Pi_{uq,V}(s)+\Pi_{uq,A}(s)\bigr].
\label{3.27}
\end{equation}
Here $V_{uq}$ are the Kobayashi--Maskawa matrix elements.  In the
massless case considered here, the vector and axial-vector hadron
correlators, $\Pi_{uq,V}$ and $\Pi_{uq,A}$ respectively, coincide, and
the function $\widetilde{R}(s)$ is equal to the ratio $R(s)$ for the
$e^+e^-$-annihilation process into hadrons.

           \begin{figure}[htb]
\centerline{ \epsfig{file=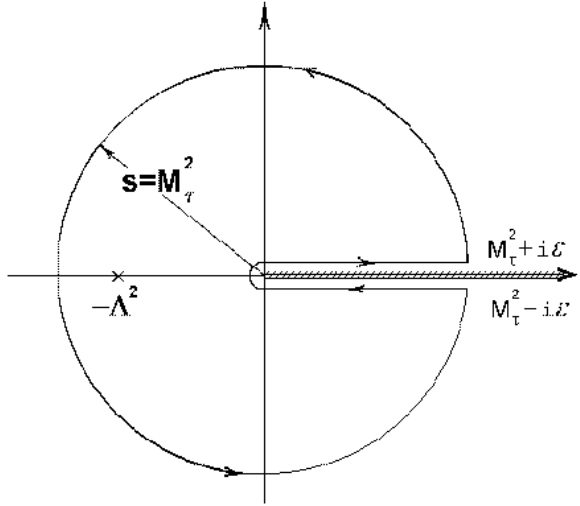,width=12.0cm}}
\caption{ {\sl
Transition to the contour representation for
$R_{\tau}$. }}
        \label{shir10}
        \end{figure}

The standard analysis of the $\tau$-decay immediately faces a
difficulty in applying the original formula~(\ref{3.26}), because the
parametrization of the function $R(s)$ by perturbative
$\bar{\alpha}_s$ with the unphysical singularities leads to
singularities in the integrand.  The way out proposed in~\cite{34}
consists in the following.  Integral~(\ref{3.26}) is represented as a
combination of integrals along the sides of the cuts in the $s$
complex plane (see Fig.~10).  By the Cauchy theorem, this integral is
then ``transformed'' into the integral along the contour
$|s|=M_{\tau}^2$.  After the integration by parts, we are left with
the contour representation of~$R_{\tau}$ involving the~$D$-function,
\begin{equation}
R_{\tau}=\frac{1}{2\pi i}
\oint_{|z|=1}\frac{dz}{z}{(1-z)}^3(1+z)D(M^2_{\tau}z).
\label{3.28}
\end{equation}

The transition from the original expression~(\ref{3.26}) to contour
representation~(\ref{3.28}) is based on certain analytic properties
of the correlator, which are violated in the standard analysis.  Thus,
the proper analytic properties ensuring the analytic approach are
important for the consistency of the inclusive $\tau$-decay
description.

We describe this process in the APT~\cite{9}.  We single out the
strong-interaction contribution~$\Delta_{\tau}$ to the
$R_{\tau}$-ratio
\begin{equation}
R_{\tau}=R_{\tau}^{(0)}(1+\Delta_{\tau}),
\label{3.29}
\end{equation}
where $R_{\tau}^{(0)}$ is a known factor including electroweak
corrections.

We express $\Delta_{\tau}$ through the effective spectral function as
\begin{equation}
\Delta_{\tau}=\frac{d_1}{\pi}\int_0^{\infty}\frac{d\sigma}{\sigma}
\rho^{\eff}(\sigma)-\frac{d_1}{\pi}\int_0^{M^2_{\tau}}
\frac{d\sigma}{\sigma}
{\left(1-\frac{\sigma}{M^2_{\tau}}\right)}^3
{\left(1+\frac{\sigma}{M^2_{\tau}}\right)}\rho^{\eff}(\sigma).
\label{3.30}
\end{equation}
Because of the universality property, the integral in the first
term can be expressed through $a_{\an}(0)$.  The spectral function in
the two-loop approximation has the form
\begin{equation}
\rho^{\eff}(\sigma)=\varrho(\sigma)+\frac{1}{\beta_0^2}\frac{d_2}{d_1}
\frac{2R(L)I(L)}{\bigl[R^2(L)+I^2(L)\bigr]^2},
\label{3.31}
\end{equation}
where the spectral density of the invariant charge
$\varrho(\sigma)$ is defined in~(\ref{2.8}) and the functions $R(L)$
and $I(L)$ are given by~(\ref{2.9}).  Inserting~(\ref{3.31})
in~(\ref{3.30}) allows us to evaluate the strong-interaction
contribution~$\Delta_{\tau}$ in terms of the scale
parameter~$\Lambda$.

Using the experimental value $R_{\tau}=3.633\pm0.031$~\cite{33}, we
obtain $\alpha(M_{\tau}^2)=0.400\pm0.026$ and the corresponding value
of the scaling parameter $\Lambda^{(3)}_{\an}=935\pm159\,$MeV.  These
values are larger than those obtained in PT using the contour
representation~\cite{35}.  The reason lies in the fact that the
nonperturbative corrections characteristic of the analytic approach
give a negative contribution to~$\Delta_{\tau}$~\cite{9,10}. Thus, to
obtain the same value~$\Delta_{\tau}$ in PT and in the analytic
approach, the ``perturbative component'' contribution of the latter
should be increased by increasing~$\Lambda$.  The inclusive
$\tau$-decay was analyzed at the three-loop APT level in~\cite{36}.
The corresponding value of~$\Lambda$ turned out to be smaller,
$\Lambda^{(3)}_{\an}=871\pm155\,$MeV.  The scheme stability of this
analysis was also shown in~\cite{36}.  It should be noted that the
quantity~$\Lambda_{\an}$ is very sensitive to the experimental value
of~$R_{\tau}$.  Thus, using $R_{\tau}=3.559\pm0.035$ \ \cite{37}, we
obtain $\Lambda^{(3)}_{\an}=640\pm127\,$MeV, which corresponds to a
considerably smaller invariant charge at the mass~$M_{\tau}$ (see
Table~1).

\sloppy
\section{The analytic approach in inelastic lep\-ton-\-had\-ron
scattering}
\label{sec4}
\fussy

In this section, we give a theoretical foundation of a possible application
of our analytic description to inelastic lepton--hadron scattering
processes.  The key point of our construction---the analytic properties of
the structure function moments with respect to ~$Q^2$---requires a certain
modification of the standard formalism, in particular, the change of the
standard Bjorken moments $M_n(Q^2)$ with the modified moments~${\cal
M}_n(Q^2)$ with respect to a new scaling variable that takes kinematic mass
dependence into account.  We start with the Jost--Lehmann integral
representation (see, for example, \S\,55 of~\cite{5}) for the Fourier image
of the corresponding matrix element.

\subsection{The Jost--Lehmann representation}

The structure functions of the inelastic lepton--hadron scattering
depend on two arguments, and the corresponding representations that
accumulate the fundamental properties of the theory (such as
relativistic invariance, spectrality, and causality) have a more
complicated form in our analysis than in representations for functions
of one variable.  Two such representations are known in the
literature.  We use the 4-dimensional integral representation proposed
by Jost and Lehmann~\cite{38} for the so-called symmetric
case.\footnote{A more general case was considered by Dyson~\cite{39},
and similar representations are therefore often called the
Jost--Lehmann--Dyson representations.}  Applications of this
representation to automodel asymptotic structure functions were
considered by Bogoliubov, Vladimirov, and Tavkhelidze~\cite{4}, some
of whose results and notation we use in what follows.  The proof of
the Jost--Lehmann representation is based on the most general
properties of the theory, such as covariance, Hermiticity,
spectrality, and causality (see~\cite{5}; some mathematical problems
related to the Jost--Lehmann--Dyson representation are also considered
in~\cite{40,41}).

For definiteness, we speak about the inelastic scattering of charged
leptons (electrons, muons) on nucleons, i.e., we consider the process
$\ell+N\rightarrow\ell+ hadrons$.  In the lowest order in the
electromagnetic coupling constant (one-photon exchange), this process
is shown in Fig.\,11, which also explains our notation.  In the
unpolarized case, the cross-section of the process is defined by the
hadronic tensor
\begin{equation}
W_{\mu\nu}(q,P)=\frac{1}{4\pi}\sum_{\sigma}\int dx\,\exp(iq\cdot x)
\left\langle P,\sigma\left|\left[J_{\mu}\left(\frac{x}{2}\right),
J_{\nu}\left(-\frac{x}{2}\right)\right]\right|P,\sigma\right\rangle
\label{4.1}
\end{equation}
constructed of the commutators of the currents, with the sum taken
over the nucleon polarizations.

            \begin{figure}[htp]
\centerline{ \epsfig{file=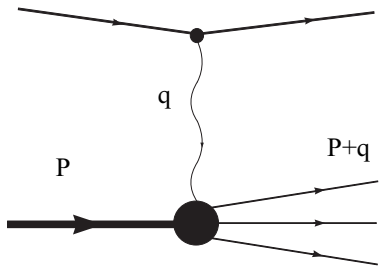,width=6.0cm}}
     \caption{  {\sl
The deep inelastic lepton--hadron scattering
diagram in the one-photon exchange approximation.
}}
\label{shir11}
          \end{figure}

Relativistic invariance and the electromagnetic current conservation
lead to the parametrization of tensor~(\ref{4.1}) in terms of two
structure functions~$w_1$ and~$w_2$,

\begin{eqnarray}\label{4.2}
W_{\mu\nu}(q,P)&=&\left(-g_{\mu\nu}+\frac{q_{\mu}q_{\nu}}{q^2}\right)
w_1(q,P) \\[0.25cm]
&+&\frac{1}{M^2}\left(p_{\mu}-\frac{P\cdot q}{q^2}q_{\mu}\right)
\left(p_{\nu}-\frac{P\cdot q}{q^2}q_{\nu}\right)w_2(q,P)\, ,
\nonumber
\end{eqnarray}
where $M=\sqrt{P^2}$ is the nucleon mass.

We now list the main properties of the functions~$w$ following from
the general principles of local QFT:

-- {\sl covariance\/} property means that the functions~$w$ depend on
    two scalar arguments, which we choose as $\nu=P\cdot q$
    and~$Q^2=-q^2$,
$$
w(q,P)=W(\nu,Q^2);
$$

-- {\sl spectrality\/} property is written as
$$
W(\nu,Q^2)=0\qquad {\rm for}\qquad\frac{Q^2}{2\nu}=x>1,
$$ where we used the dimensionless Bjorken variable, which in the
physical domain of the process for $(q+P)^2>M^2$ is kinematically
restricted by the interval~$0<x<1$;

-- the structure function parametrizes the scattering cross-section
   and is real (the {\sl reality\/} property),
$$
W(\nu,Q^2)=W^*(\nu,Q^2);
$$

-- Hermiticity of the current operator leads to the (anti-){\sl
    symmetry\/} property
$$
W(-\nu,Q^2)=-W(\nu,Q^2);
$$

-- the vanishing of the commutator of currents at space-like intervals
    because of the local commutativity of currents gives the {\sl
    causality\/} condition
$$ \int\frac{dq}{(2\pi)^4}\exp(-iqz)W(q,P)=0\qquad {\rm for}\qquad z^2<0. $$
For the function $W(\nu,Q^2)$ satisfying all these conditions, there exists
a real moderately growing distribution $\psi({\bf u},\lambda^2)$ such that
the Jost--Lehmann integral representation holds; in the nucleon rest frame,
this can be written as~\cite{4}
\begin{equation}
W(\nu,Q^2)=\varepsilon(q_0)\int d{\bf u}\,d\lambda^2\,
\delta\bigl[q_0^2-(M {\bf u}-{\bf q})^2-\lambda^2\bigr]
\psi({\bf u},\lambda^2),
\label{4.3}
\end{equation}
with the function $\psi({\bf u},\lambda^2)$ supported on the set
$$
\rho=|{\bf u}|\leq1,\qquad
\lambda^2\geq M^2\bigl(1-\sqrt{1-\rho^2}\,\bigr)^2.
$$

For the process under consideration, the physical values of~$\nu$
and~$Q^2$ are positive.  We, thus, can neglect the factor
$\varepsilon(q_0)= \varepsilon(\nu)$ and keep the same notation
for~$W(\nu,Q^2)$.  Taking into account that the weight function
$\psi({\bf u},\lambda^2)=\psi(\rho,\lambda^2)$ is radial-symmetric,
as follows from covariance~\cite{4}, we write the Jost--Lehmann
representation for~$W$ in the covariant form,
\begin{eqnarray}\label{4.4}
W(\nu,Q^2)&=&\int_0^1d\rho\,\rho^2
\int_{\lambda_{\min}^2}^{\infty}d\lambda^2\int_{-1}^{1}dz\\[0.25cm]
&\times&\delta\bigl(Q^2+M^2\rho^2+\lambda^2-
2z\rho\sqrt{\nu^2+M^2Q^2}\,\bigr)\psi(\rho,\lambda^2)\,,
\nonumber
\end{eqnarray}
where
\begin{equation}
\lambda_{\min}^2=M^2\bigl(1-\sqrt{1-\rho^2}\,\bigr)^2.
\label{4.5}
\end{equation}

\subsection{Analytic moments of the structure functions}

As follows from representation~(\ref{4.4}), a natural scaling
variable is given by
\begin{equation}
s=\frac{1}{2}\sqrt{\frac{Q^2(Q^2+4M^2)}{\nu^2+M^2Q^2}}\,=
x\sqrt{\frac{Q^2+4M^2}{Q^2+4M^2x^2}}\,,
\label{4.6}
\end{equation}
which accumulates the root structure determined by the
$\delta$-function argument.  At the same time, in the physical region
of the process, the $s$ variable changes in the same way as the
Bjorken variable~$x$, i.e., from zero to one.  The variable~$s$ bears
a dependence on the mass of the target (the nucleon) and is different
from both the Bjorken variable and the Nachtmann variable~\cite{42}
$$
\xi=\frac{2 x}{1+\sqrt{1+x^2 4M^2/Q^2}}
$$ that is sometimes used in the kinematical account of mass effects
of inelastic scattering processes.  However, only the~$s$ variable
leads to the moments that have the analytic properties in~$Q^2$ that
we need.

To establish these properties, we integrate over~$z$ in~(\ref{4.4})
as
\begin{eqnarray}
\label{4.7}
W(\nu,Q^2)&=&\frac{s}{Q^2\sqrt{1+4M^2/Q^2}}\int_0^1d\rho\,\rho
\int_{\lambda_{\min}^2}^{\infty}d\lambda^2      \\[0.25cm]
&\times&\theta\left[Q^2\rho^2-s\frac{(Q^2+M^2\rho^2+\lambda^2)}
{\sqrt{1+4M^2/Q^2}}\right]\,\psi(\rho,\lambda^2).
\nonumber
\end{eqnarray}

We now define the modified $s$-moments of the structure functions
(cf.~\cite{43})
\begin{equation}
{\cal M}_n(Q^2)=\frac{1}{{(1+4M^2/Q^2)}^{(n-1)/2}}
\int\limits_0^{1}  d s s^{n-2} W(\nu,Q^2) \, .
\label{4.8}
\end{equation}
Inserting $W(\nu,Q^2)$ as given by~(\ref{4.7}), we obtain
$$
{\cal M}_n(Q^2)=\frac{(Q^2)^{n-1}}{n}\int_0^1d\rho\,\rho^{n+1}
\int_{0}^{\infty}d\sigma
\frac{\theta(\sigma-\sigma_{\min})}{(Q^2+\sigma)^{n}}
\psi(\rho,\sigma-M^2\rho^2),
$$
where $\sigma=\lambda^2+M^2\rho^2$ and
$\sigma_{\min}=2M^2\bigl(1-\sqrt{1-\rho^2}\,\bigr)$.

Introducing the weight function
\begin{equation}
m_n(\sigma)=\frac{1}{n}\int_0^1d\rho\,\rho^{n+1}
\theta(\sigma-\sigma_{\min})\,\psi(\rho,\sigma-M^2\rho^2),
\label{4.9}
\end{equation}
we obtain the representation for the $s$-moments
\begin{equation}
{\cal M}_n(Q^2)=(Q^2)^{n-1}\int_0^{\infty} d\sigma
\frac{m_n(\sigma)}{(\sigma+Q^2)^n},
\label{4.10}
\end{equation}
which implies the analyticity of ${\cal M}_n(Q^2)$ in the complex
$Q^2$ plane cut along the negative semi-axis, i.e., the
K\"all\'en--Lehmann type analyticity.

In~\cite{44}, the Deser--Gilbert--Sudarshan integral
representation~\cite{45} was used to arrive at a similar statement
regarding the analyticity of the K\"all\'en--Lehmann type for the
$x$-moments.  However the status of this representation in QFT is less
clear, since it cannot be obtained starting with only the basic
principles of the theory.

The relation between moments~(\ref{4.8}) and the standard Bjorken
moments
\begin{equation}
M_n(Q^2)=\int_0^1dx\,x^{n-2}W(\nu,Q^2)
\label{4.11}
\end{equation}
can be expressed by
\begin{eqnarray}
{\cal M}_n(Q^2)&=&
\frac{1}{\Gamma[(n+1)/2]}\,\sum\limits_{k=0}^\infty
\frac{\Gamma[k+(n+1)/2]}{k!}\left(-\frac{4M^2}{Q^2}\right)^k M_{n+2k}(Q^2)
\, ,    \nonumber  \\
M_n(Q^2)&=&\frac{1}{\Gamma[(n+1)/2]}\,\sum\limits_{k=0}^\infty
\frac{\Gamma[k+(n+1)/2]}{k!}\left(\frac{4M^2}{Q^2}\right)^k
{\cal M}_{n+2k}(Q^2) \, .    \nonumber
\end{eqnarray}

In the asymptotic domain corresponding to large values of the
transferred momentum~$Q^2$, where power corrections of the form
$1/(Q^2)^n$ can be neglected, the $x$-, $s$-, and~$\xi$-moments are
identical.  Outside the asymptotic domain, on the other hand, where it
comes to studying the contribution of higher twists, the difference
between these definitions of moments must be taken into account.

\subsection{Dispersion relation and the operator product expansion}

To establish the relation with the operator product expansion, we
start with the Jost--Lehmann representation and obtain a dispersion
relation for the forward Compton scattering amplitude with respect to
the new variable~(\ref{4.6}).  We write the matrix element of the
process corresponding to representation~(\ref{4.4}) as
\begin{eqnarray}\label{4.12}
T(\nu,Q^2)&=&\frac{1}{\pi}\int_0^1d\rho\,\rho^2
\int_{\lambda_{\min}^2}^{\infty}d\lambda^2 \int_{-1}^{1}\,dz\\[0.25cm]
&\times&\frac{\psi(\rho,\lambda^2)}
{Q^2+M^2\rho^2+\lambda^2-2z\rho\sqrt{\nu^2+M^2Q^2}\,-i\epsilon}.
\nonumber
\end{eqnarray}

In the complex $\nu^2$ plane, the function $T(\nu,Q^2)$ has a branch
cut along the positive part of the real axis starting
at~$\nu^2_{\min}$ defined by the condition
$$
\sqrt{\nu^2_{\min}+M^2Q^2}\,=\min_{\{\lambda,\rho,z\}}
\left|\frac{Q^2+M^2\rho^2+\lambda^2}{2z\rho}\right|. $$

Recalling~(\ref{4.5}) and the range of the integration
variable~$z$ in~(\ref{4.12}), we can simplify this to
\begin{equation}
\sqrt{\nu^2_{\min}+M^2Q^2}\,=\min_{\{\rho\}}
\frac{Q^2+2M^2\bigl(1-\sqrt{1-\rho^2}\,\bigr)}{2\rho},
\label{4.13}
\end{equation}
which leads to $\nu^2_{\min}=(Q^2/2)^2$.  Thus, the sought
dispersion relation has the form
\begin{equation}
T(\nu,Q^2)=\frac{1}{\pi}\int_{Q^4/4}^{\infty}
\frac{d\nu_1^2}{\nu_1^2-\nu^2-i\epsilon}W(\nu_1,Q^2).
\label{4.14}
\end{equation}

We note that in terms of the Bjorken variable~$x$,
relation~(\ref{4.14}) is represented as
\begin{equation}
T(\nu,Q^2)=\frac{2}{\pi}\int_0^1\frac{dx_1}{x_1}
\frac{1}{1-(x_1/x)^2}\,W(\nu_1,Q^2).
\label{4.15}
\end{equation}
This expression determines simple properties of the amplitude
$T(x,Q^2)$ in the complex $x$-plane and is convenient in the operator
product expansion.

In considering consequences of the Jost--Lehmann representation, as
noted above, the natural scaling variable is given by~$s$.  In this
case, there arises a similar structure of the dispersion
integral~\footnote{We note that in using other scaling variables, for
example, the Nachtmann one, this structure can be destroyed.}
\begin{equation}
T(\nu,Q^2)=\frac{2}{\pi}\int_0^{1}\frac{ds_1}{s_1}\frac{1}{1-(s_1/s)^2}
\,W(\nu_1,Q^2);\qquad
\nu^2=Q^2\left[\frac{Q^2+4M^2}{4s^2}-M^2\right].
\label{4.16}
\end{equation}

The identity between the structures of the dispersion relations with
respect to the variables~$x$ and~$s$ allows us to establish the
relation of analytic moments~(\ref{4.8}) to the operator product
expansions of currents used in finding the~$Q^2$-evolution of the
structure functions of the moments.  The moments in Eq.~(\ref{4.11})
correspond to the case where only the Lorentz structures of the form
$P_{\mu_1}\dots P_{\mu_n}$ are taken into account in matrix elements
of the operator~$\bigl\langle
P|\widehat{O}_{\mu_1\dots\mu_n}|P\bigr\rangle$.  Then the application
of the operator product expansion for the Compton amplitude leads to
the expansion in powers of $(q\cdot P)/Q^2$, i.e., to the expansion in
the inverse powers of~$x$.  A similar expansion in the inverse powers
of~$x$ can also be done in dispersion integral~(\ref{4.15}).  The
coefficients are then determined by the $x$-moments.  Comparing the
two power series gives the sought relation between the $x$-moments and
the operator product expansion.

In the general case, the symmetric matrix element $\bigl\langle
P|\widehat{O}_{\mu_1\dots\mu_n}|P\bigr\rangle$ contains the Lorentz
structures given by $\{P_{\mu_1}\dots P_{\mu_n}\}$, \
$M^2g_{\mu_i\mu_j}\{P_{\mu_1}\dots P_{\mu_{n-2}}\}$, etc.  The moments
with respect to the~$\xi$ variable correspond to choosing the operator
basis where the expansion goes over traceless tensors, i.e., such that
the contraction of $g_{\mu_i\mu_j}$ with $\bigl\langle
P|\widehat{O}_{\mu_1\dots\mu_n}|P\bigr\rangle$ vanishes for any two
indices.  It is then obvious that the Lorentz structure of the matrix
element $\bigl\langle P|\widehat{O}_{\mu_1\dots\mu_n}|P\bigr\rangle$
is fixed unambiguously.

Dispersion representation~(\ref{4.16}) allows us to expand the
Compton amplitude in the inverse powers of~$s$.  If the operator basis
is chosen such that an arbitrary contraction of the
tensor~$\bigl\langle P|\widehat{O}_{\mu_1\dots\mu_n}|P\bigr\rangle$
with the nucleon momentum~$P_{\mu_i}$ vanishes, then the operator
product expansion leads to a power series for the forward Compton
scattering amplitude with the expansion parameter $q_{\mu}
q_{\nu}(P_{\mu} P_{\nu}-g_{\mu\mu}P^2)/(q^2)^2$, which corresponds to
expanding dispersion integral~(\ref{4.16}) in powers of~$1/s^2$.  We
thus arrive at the relation between the analytic~$s$-moments and the
operator product expansion.  We stress that the orthogonality
requirement of the symmetric tensor $\bigl\langle
P|\widehat{O}_{\mu_1\dots\mu_n}|P\bigr\rangle$ to the nucleon
momentum~$P_{\mu_i}$ determines its Lorentz structure unambiguously.

\section{Conclusions}

We considered the analytic formulation of QCD, where the analyticized
RG-solutions for the invariant coupling functions, the Green's
functions, and the matrix elements are free of unphysical
singularities.  An important property of this formulation that we
found is the stability of the analytic invariant charge with respect
to higher-loop correction in all of the~$Q^2$ range.  The key point
here is the existence of the universal limiting value
$\bar{\alpha}_{\an}(0)=4\pi/\beta_0$ that is invariant with respect to
multiloop corrections.  This constant is independent of the
$\Lambda_{\rm {QCD}}$ parameter and is determined only by the general
symmetry properties of the Lagrangian.  Therefore, the family of
curves $\bar{\alpha}_{\an}(Q^2/\Lambda^2)$ for different values of
the~$\Lambda$ parameter is a bundle with the common point
$\bar{\alpha}_{\an}(0)=4\pi/\beta_0$ (this picture is independent of
the number of loops).

The invariant analytic formulation essentially modifies the behavior
of $\bar{\alpha}_{\an}(x)$ in the IR region by making it stable with
respect to higher-loop corrections.  The two-loop approximation
differs from the one-loop one by no more than $\simeq10\,\%$ in the
small-$Q^2$ domain, and the three-loop approximation differs from the
two-loop one by only $\simeq1\,\%$.  This is radically different from
the situation encountered in the standard renormalization-group PT,
which is characterized by strong instability with respect to the next
loop corrections in the domain of small~$Q^2\simeq\Lambda^2$.  We note
also that maintaining the proper analytic properties with respect
to~$Q^2$ is essential for a self-consistent definition of the
effective coupling function in the time-like region~\cite{15}.  In
describing the concrete processes, for example the inclusive
$\tau$-lepton decay, a consistent analysis is possible~\cite{9} only
provided the above analytic properties hold.

There are at least two possibilities to describe physical quantities
in the new approach framework.  The simplest one consists in replacing
$\bar{\alpha}_s(x)\to\bar{\alpha}_{\an}(x)$ in the explicit
expressions for the observables ``processed'' by the RG method, or
more precisely, for the related quantities defined in the space-like
region of the~$Q^2$ variable.

We take another possibility, however.  For the quantities similar to
the Adler $D(Q^2)$-function that are represented by the PT power
series, according to a special convention, the analyticization
procedure is applied to each power of $\bar{\alpha}_s(Q^2)$
separately.  This leads to a new non-power-series expansion, in which
the powers of $\bar{\alpha}_s(Q^2)$ are replaced with new nonsingular
functions of~$A_n(Q^2)$.  We call this algorithm, which was first
proposed in~\cite{9}, the APT.  Applying this algorithm to analyze the
amplitudes of the processes like the $e^+e^-$-annihilation into hadrons
and the inclusive $\tau$-decay, and also of the sum rules for the
inelastic lepton--hadron scattering, we see that in addition to
possessing loop stability, the APT results are much less sensitive to
the choice of the renormalization scheme than in the standard
approach.  In other words, the three-loop APT level practically
insures both the loop saturation and the scheme invariance of the
relevant physical quantities in the entire energy or momentum range.

It appears that by accounting for the additional information about the
proper analytic properties, the first terms of the APT
non-power-series expansion already give sufficiently good
approximation to the sum of the whole series.  We recall here the
analogy with summing up the perturbative expansions with the
additional information on the behavior of the remote PT series terms
taken into account~\cite{46}.  In that case, it also turned out that
the expression for the approximated function given by the first
several terms of the loop expansion was practically unchanged by
higher corrections.

In this work, we considered also the structure functions of the
inelastic lepton--hadron scattering, which are more complicated
objects than the two-point functions, which are in one way or another
related with the K\"all\'en--Lehmann representation.  For these
functions, the general quantum field theory principles, including
covariance, Hermiticity, spectrality, and causality, are expressed by
the Jost--Lehmann--Dyson integral representation.  In using the
analytic approach to define the $Q^2$-evolution, it was convenient to
introduce the moments ${\cal M}_n(Q^2)$ of the structure functions
corresponding to the special scaling variable~(\ref{4.6}).  It is
these moments, rather than the Bjorken or Nachtmann ones, that
exhibits simple analytic properties with respect to~$Q^2$.  In this
work, we found the relation of the new analytic moments
${\cal M}_n(Q^2)$ to the operator product expansion, where the tensor
structure of the matrix elements of operators with respect to the
nucleon states must be fixed according to the condition that they be
orthogonal to the nucleon momentum.

\vspace{0.5cm}
\centerline{\bf Acknowledgments}
\vspace{0.2cm}

It is a pleasure to thank V.~S.~Vladimirov, A.~V.~Efre\-mov,
B.~I.~Zav'yalov, V.~A.~Meshcheryakov, S.~V.~Mikhailov, K.~A.~Milton,
O.~V.~Teryaev, and O.~P.~Solovtsova for useful discussions of the results.
We are also grateful to the Russian Foundation for Basic Research (Grants
No.~96-15-96030 and No.~99-01-00091) and INTAS (Grant No.~96-0842) for
support.


\begin{thebibliography}{99}

\bibitem{1}
N.N.~Bogoliubov and D.V.~Shirkov,
{\it Dokl. Akad. Nauk SSSR} {\bf 103}, 203 (1955); 391.

\bibitem{2}
N.~N.~Bogoliubov and D.~V.~Shirkov,
{\it JETP} {\bf 30}, 77 (1956);
{\it Nuovo Cim.} {\bf 3}, 845 (1956).

\bibitem{3}
N.~N.~Bogolyubov, A.~A.~Logunov, and D.~V.~Shirkov,
{\it JETP} {\bf 37}, 805 (1959).

\bibitem{4}
N.~N.~Bogolyubov, V.~S.~Vladimirov, and A.~N.~Tavkhelidze,
{\it Theor. Math. Phys.} {\bf 12}, 3 (1972); 305.

\bibitem{5}
N.~N.~Bogoliubov and D.~V.~Shirkov,
{\it Introduction to the Theory of Quantum Fields
(Chap. ``Desipersion Relations'')}
[{\sl in Russian}], Nauka, Moscow (1973, 1976, 1986);
{\sl English transl.}: Wiley, New York (1959, 1980).

\bibitem{6}
D.~V.~Shirkov and I.~L.~Solovtsov,
{\it JINR Rap. Comm.} {\bf 2}~[76], 5 (1996);
``Analytic QCD running coupling with finite IR behavior and universal
$\bar{\alpha}_s (0)$ value", hep-ph/9604363.

\bibitem{7}
D.~V.~Shirkov and I.~L.~Solovtsov,
{\it Phys. Rev. Lett.} {\bf 79}, 1209 (1997).

\bibitem{8}
I.~L.~Solovtsov and D.~V.~Shirkov,
{\it Phys. Lett.~B} {\bf 442}, 344 (1998).

\bibitem{9}
K.~A.~Milton, I.~L.~Solovtsov, and O.~P.~Solovtsova,
{\it Phys. Lett.~B} {\bf 415}, 104 (1997).

\bibitem{10}
O.~P.~Solovtsova,
{\it JETP Lett.} {\bf 64}, 664 (1996).

\bibitem{11}
K.~A.~Milton, I.~L.~Solovtsov, and O.~P.~Solovtsova,
``Analytic Perturbative Approach to QCD",
Talk given at the XXIX Int. Conference on HEP,
Vancouver, B.~C., Canada,
July 23-29, 1998 (to be published in the Proceed.);
Preprint OKHEP-98-06, Oklahoma, Oklahoma Univ. (1998);
hep-ph/9808457.

\bibitem{12}
D.~V.~Shirkov,
{\it Nucl. Phys. B (Proc. Suppl.)}  {\bf 64}, 106 (1998), hep-ph/9708480;\\
{\it Theor. Math. Phys.} {\bf 19}, 438 (1999);
``Renormalization group, causality, and nonpower perturbation
expansion in QFT"; Preprint E2-98-311, JINR, Dubna (1998);
hep-th/9810246.

\bibitem{13}
K.~A.~Milton, I.~L.~Solovtsov, and O.~P.~Solovtsova,
{\it Phys. Lett.~B} {\bf 439}, 421 (1998).

\bibitem{14}
K.~A.~Milton, I.~L.~Solovtsov, and O.~P.~Solovtsova,
{\it Phys. Rev.~D} {\bf 60} 016001 (1999);
Preprint OKHEP-98-07, Oklahoma, Oklahoma Univ. (1998);
hep-ph/9809513.

\bibitem{15}
K.~A.~Milton and I.~L.~Solovtsov,
{\it  Phys. Rev.~D} {\bf 55}, 5295 (1997).

\bibitem{16}
K.~A.~Milton and O.~P.~Solovtsova,
{\it  Phys. Rev.~D} {\bf 57}, 5402 (1998).

\bibitem{17}
I.~F.~Ginzburg and D.~V.~Shirkov,
{\it JETP} {\bf 22}, 234 (1966).

\bibitem{18}
D.~V.~Shirkov,
{\it Nucl. Phys.~B} {\bf 332}, 425 (1990).

\bibitem{19}
D.~V.~Shirkov,
{\it Lett. Math. Phys.} {\bf 1}, 179 (1976).

\bibitem{20}
B.~A.~Magradze,
``The gluon propagator in analytic perturbation theory",
Talk given at 10th Intern. Seminar on
High-Energy Physics (Quarks 98), Suzdal, Russia, 18-24 May, 1998;
Preprint G-TMI-98-08-01, Tbilisi, TMI (1998);
hep-ph/9808247.

\bibitem{21}
E.~Gardi, G.~Grunberg, and M.~Karliner,
``Can the QCD running coupling have a
causal analyticity structure?",
Preprint TAUP-2503-98, Paris, TAUP (1998);
hep-ph/9806462.

\bibitem{22}
A.~C.~Mattingly and P.~M.~Stevenson,
{\it Phys. Rev.~D} {\bf 49}, 437 (1994).

\bibitem{23}
Yu.~L.~Dokshitzer, V.~A.~Khoze, and S.~I.~Troyan,
{\it Phys. Rev.~D} {\bf 53}, 89 (1996).

\bibitem{24}
E.~C.~Poggio, H.~R.~Quinn, and S.~Weinberg,
{\it Phys. Rev.~D} {\bf 13}, 1958 (1976).

\bibitem{25}
P.~M.~Stevenson,
{\it Phys. Rev.~D} {\bf 23}, 2916 (1981).

\bibitem{26}
S.~G.~Gorishny, A.~L.~Kataev, and S.~A.~Larin,
{\it Phys. Lett.~B} {\bf 259}, 144 (1991).

\bibitem{27}
F.~Jegerlehner,
{\it Nucl. Phys.~C~(Proc. Suppl.)} {\bf 51}, 131 (1996);
``Hadronic vacuum polarization contribution to $g-2$ of the leptons and
$\alpha (M_z)$", DESY Preprint 96-121, Hamburg, DESY (1996); hep-ph/9606484.

\bibitem{28}
S.~Eidelman, F.~Jegerlehner, A.~L.~Kataev, and O.~Veretin,
{\it Phy. Lett.~B} {\bf 454}, 369 (1999);
``Testing nonperturbative strong interaction effects via the Adler
function", Preprint DESY 98-206,
Hamburg, DESY (1998); hep-ph/9812521.

\bibitem{29}
J.~Chyla, A.~L.~Kataev, and S.~A.~Larin,
{\it Phys. Lett.~B} {\bf 261}, 269 (1991).

\bibitem{30}
P.~A.~R\c{a}czka and A.~Szymacha,
{\it Phys. Rev.~D} {\bf 54}, 3073 (1996).

\bibitem{31}
W.~Celmaster and R.~J.~Gonsalves,
{\it Phys. Rev.~D} {\bf 20}, 1420 (1979).

\bibitem{32}
P.~A.~R\c{a}czka,
{\it Z.~Phys.~C} {\bf 65}, 481 (1995).

\bibitem{33}

Particle Data Group,
{\it Phys. Rev. D} {\bf 54}, 1 (1996);
{\it Eur.~Phys.~J.~C} {\bf 3}, 1 (1998).

\bibitem{34}
E.~Braaten,
{\it Phys. Rev. Lett.} {\bf 60}, 1606 (1988);
{\it Phys. Rev.~D} {\bf 39}, 1458 (1989).

\bibitem{35}
E.~Braaten, S.~Narison, and A.~Pich,
{\it Nucl. Phys.~B} {\bf 373}, 581 (1992).

\bibitem{36}
K.~A.~Milton, I.~L.~Solovtsov, and V.~I.~Yasnov,
``Analytic perturbation theory and renormalization scheme dependence
in $\tau$-decay",
Preprint  OKHEP-98-01, Oklahoma, Oklahoma Univ. (1998);
hep-ph/9802262.

\bibitem{37}
T.~Coan et al. {\rm (}CLEO Collaboration\/{\rm)},
{\it Phys. Lett.~B} {\bf 356}, 580 (1996).

\bibitem{38}
R.~Jost and H.~Lehmann,
{\it  Nuovo Cim.} {\bf 5}, 1598 (1957).

\bibitem{39}
F.~J.~Dyson,
{\it Phys. Rev.} {\bf 110}, 1460 (1958).

\bibitem{40}
N.~N.~Bogoliubov, A.~A.~Logunov, A.~I.~Oksak, and I.~T.~Todorov,
{\it General Principles of Quantum Field Theory}
[{\sl in Russian}], Nauka, Moscow (1987);
{\sl English transl.}: Dordrecht, Kluwer (1990).

\bibitem{41}
V.~S.~Vladimirov, Yu.~N.~Drozhzhinov, and B.~I.~Zav'yalov,
{\it Tauberian Theorems for General Functions}
[{\sl in Russian}], Nauka, Moscow (1986);
{\sl English transl.}: Dordrecht, Kluwer (1988).

\bibitem{42}
O.~Nachtmann,
{\it Nucl. Phys.~B} {\bf 63}, 237 (1973).

\bibitem{43}
B.~Geyer, D.~Robaschik, and E.~Wieczorek,
{\it Fortschr. Phys.} {\bf 27}, 75 (1979);
{\it  Fiz. \`Elementar. Chastits i Atom. Yadra} {\bf 11}, 132 (1980).

\bibitem{44}
W.~Wetzel,
{\it Nucl. Phys.~B} {\bf 139}, 170 (1978).

\bibitem{45}
S.~Deser, W.~Gilbert, and E.~C.~S.~Sudarshan,
{\it Phys. Rev.} {\bf 117}, 266 (1960).

\bibitem{46}
D.~I.~Kazakov, O.~V.~Tarasov, and D.~V.~Shirkov,
{\it Theor. Math. Phys.} {\bf 38}, 9 (1979);
D.~I.~Kazakov and D.~V.~Shirkov,
{\it Fortschr. Phys.} {\bf 28}, 465 (1980).

\end{thebibliography}
\end{document}